\documentclass[pdflatex,sn-nature,oneside]{sn-jnl}

\usepackage{graphicx}%
\usepackage{multirow}%
\usepackage{amsmath,amssymb,amsfonts}%
\usepackage{amsthm}%
\usepackage{mathrsfs}%
\usepackage[title]{appendix}%
\usepackage{xcolor}%
\usepackage{textcomp}%
\usepackage{manyfoot}%
\usepackage{booktabs}%
\usepackage{algorithm}%
\usepackage{algorithmicx}%
\usepackage{algpseudocode}%
\usepackage{listings}%
\usepackage{newtxmath}
\usepackage{geometry}
\usepackage[normalem]{ulem}

\theoremstyle{thmstyleone}%
\theoremstyle{thmstyletwo}%

\theoremstyle{thmstylethree}%

\definecolor{revisioncolor}{HTML}{3370BD}

\begin{document}

\title[Article Title]{Room-temperature hybrid 2D-3D quantum spin system for enhanced magnetic sensing and many-body dynamics}


\author[1,2]{\fnm{Haoyu} \sur{Sun}}
\equalcont{These authors contributed equally to this work.}

\author[1,2]{\fnm{Pei} \sur{Yu}}
\equalcont{These authors contributed equally to this work.}

\author[1,2,3]{\fnm{Xu} \sur{Zhou}}
\equalcont{These authors contributed equally to this work.}

\author[1,2]{\fnm{Xiangyu} \sur{Ye}}

\author*[1,2]{\fnm{Mengqi} \sur{Wang}}\email{mqw@ustc.edu.cn} 

\author[1,2]{\fnm{Zhaoxin} \sur{Liu}}

\author[1,2]{\fnm{Yuhang} \sur{Guo}}

\author[1,2]{\fnm{Wenzhao} \sur{Liu}}

\author[1,2]{\fnm{You} \sur{Huang}}

\author[1,2,3]{\fnm{Pengfei} \sur{Wang}}

\author[1,2,3]{\fnm{Fazhan} \sur{Shi}}

\author[1,2,3]{\fnm{Kangwei} \sur{Xia}}

\author*[1,2,3]{and \fnm{Ya} \sur{Wang}}\email{ywustc@ustc.edu.cn} 

\affil[1]{\orgdiv{CAS Key Laboratory of Microscale Magnetic Resonance and School of Physical Sciences}, \orgname{University of Science and Technology of China}, \orgaddress{\city{Hefei} \postcode{230026}, \country{China}}}

\affil[2]{\orgdiv{Anhui Province Key Laboratory of Scientific Instrument Development and Application}, \orgname{University of Science and Technology of China}, \orgaddress{\city{Hefei} \postcode{230026}, \country{China}}}

\affil[3]{\orgdiv{Hefei National Laboratory}, \orgname{University of Science and Technology of China}, \orgaddress{\city{Hefei} \postcode{230088}, \country{China}}}



\abstract{Advances in hybrid quantum systems and their precise control are pivotal for developing advanced quantum technologies. Two-dimensional (2D) materials with optically accessible spin defects have emerged as a promising platform for building integrated quantum spin systems due to their exceptional flexibility and scalability. However, experimentally realizing such systems and demonstrating their superiority remains challenging. Here, we present a hybrid spin system operating under ambient conditions, integrating boron vacancy (V${\rm{_{B}^{-}}}$) spins in 2D hexagonal boron nitride flakes with a single nitrogen vacancy (NV) center in 3D single-crystal diamonds. This combined system achieves full controllability and exhibits enhanced performance for nanoscale magnetic sensing, including an improved dynamic range. Moreover, we investigate the rich many-body spin dynamics within the hybrid system, which enables us to estimate the concentration of V${\rm{_{B}^{-}}}$ spins. This work provides a critical foundation for advancing the development of 2D-3D integrated quantum spin systems.}


\maketitle

\section*{Introduction}\label{sec1}

Hybrid quantum systems have emerged as versatile platforms, thanks to advancements in engineering and precise quantum control, which can be utilized in quantum applications across quantum networks \cite{wilk2007single,kimble2008quantum,dong2015optomechanical,reiserer2015cavity}, quantum simulation \cite{cai2013large,georgescu2014quantum}, and quantum sensing \cite{ajoy2015atomic,zaiser2016enhancing}. Several hybrid systems have been intensively developed, including atom-cavity systems \cite{wilk2007single,reiserer2015cavity}, superconducting circuit-spin coupling \cite{xiang2013hybrid,clerk2020hybrid}, and optomechanical systems \cite{dong2015optomechanical,kurizki2015quantum}, both to explore fundamental physics and to expand the array of systems. However, these systems require specialized engineering and extreme conditions to maintain their quantum properties, hampering their applications in quantum technologies. To date, developing superior hybrid quantum systems that can work reliably at room temperature remains an outstanding challenge.

Recently, two-dimensional (2D) materials and their associated heterostructures have attracted considerable attention \cite{geim2013van,novoselov20162d,wang2021van}, both in the development of high-performance optoelectronic devices \cite{liu2016van,liang2020van} and in the observation of unique physical phenomena or properties \cite{geim2013van,hunt2013massive,spanton2018observation,jin2019observation}. Moreover, the exceptional flexibility and scalability of 2D materials, combined with optically accessible individual spin defects within them, offer new opportunities for constructing hybrid or scalable quantum systems. Recent advances in materials engineering have thus focused on creating optically detectable single spin defects, such as the negatively charged boron vacancy (V${\rm{_{B}^{-}}}$) spins \cite{gottscholl2020initialization} and carbon-related defects \cite{chejanovsky2021single,mendelson2021identifying,stern2022room,guo2023coherent,stern2024quantum}. While ensemble V${\rm{_{B}^{-}}}$ spins have been extensity investigated \cite{vaidya2023quantum}, their integration into functional hybrid quantum systems remains underexplored, leaving room to leverage their potential for quantum technologies. 

In this work, we present a hybrid spin system that operates under ambient conditions by constructing a 2D-3D heterostructure. This system consists of a V${\rm{_{B}^{-}}}$ spin ensemble in thin 2D hexagonal boron nitride flakes and a single nitrogen vacancy (NV) center in three-dimensional (3D) single-crystal diamonds. By integrating these two distinct spin systems, we achieve separate optical readout and independent microwave control of NV and V${\rm{_{B}^{-}}}$ spins, enabling hybrid nanoscale magnetic sensing with an enhanced dynamic range. Furthermore, by characterizing their interaction strengths, we estimate the concentration of V${\rm{_{B}^{-}}}$ spins and observe the resulting coherent spin dynamics. These results establish 2D-3D hybrid spin systems as a promising platform for quantum sensing and spin-based quantum technologies.

\section*{Results}\label{sec2}

\subsection*{Optical and microwave addressability of NV and V${\rm{_{B}^{-}}}$ spins}\label{subsec1}

\begin{figure}[htbp]
\centering
\includegraphics[width=1\textwidth]{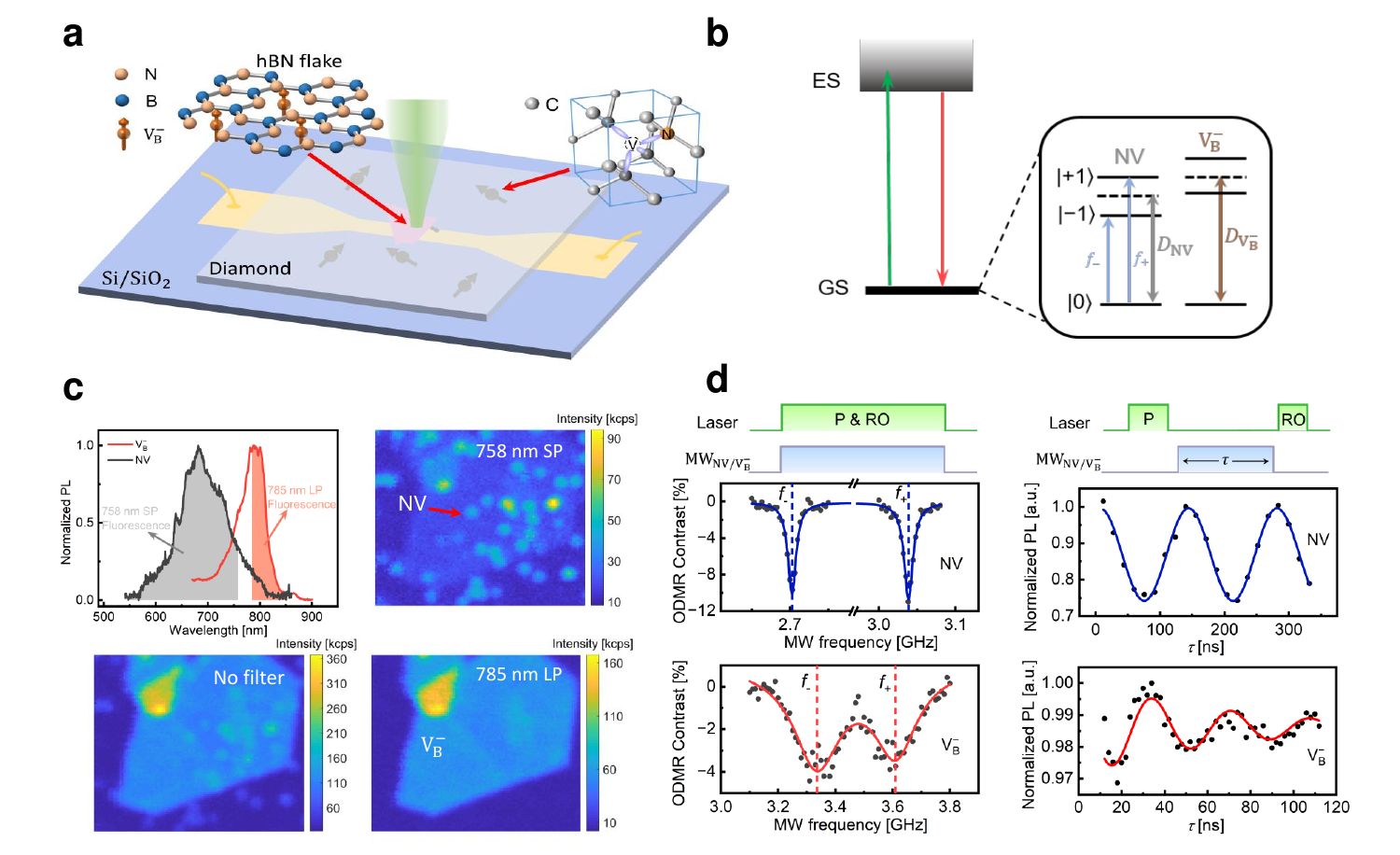}
\caption{\textbf{Optical and microwave addressability of NV and V${\rm{_{B}^{-}}}$ spins.} 
\textbf{a} Schematic of the experimental configuration for spin manipulation and the atomic structures of NV and V${\rm{_{B}^{-}}}$ centers. The diamond with the transferred hBN flakes is placed upside down onto a gold film microwave stripline. \textbf{b} Simplified energy level diagram of NV and V${\rm{_{B}^{-}}}$ spins, illustrating optical transitions between the ground state (GS) and the excited state (ES). $D_{\rm{NV}}$ and $D_{\rm{V_{B}^{-}}}$ denote the zero-field splitting parameters. Transition frequencies: $f_-$($m_{s} = -1 \leftrightarrow m_{s} = 0$) and $f_+$($m_{s} = +1 \leftrightarrow m_{s} = 0$). \textbf{c} Top left: Normalized photoluminescence spectra of NV and V${\rm{_{B}^{-}}}$ centers excited with 532 nm laser at room temperature. The shaded regions indicate the fluorescence collected using short-pass (SP) or long-pass (LP) filter. Bottom left $\&$ right: Confocal PL images of NV and V${\rm{_{B}^{-}}}$ spins obtained without a filter (bottom left), with a 758 nm SP filter (top right), and with a 785 nm LP filter (bottom right). \textbf{d} Continuous-wave (CW) ODMR spectra and Rabi oscillations of NV (blue) and V${\rm{_{B}^{-}}}$ (red) spins. Left: CW ODMR spectra obtained under simultaneous MW driving, optical pumping, and readout. Right: Rabi oscillations for the $m_{s} = -1 \leftrightarrow m_{s} = 0$ transition, measured using a variable-length MW pulse following optical excitation.}\label{fig.1}
\end{figure}

To construct a hybrid 2D-3D quantum spin system, we exfoliated hBN flakes using adhesive tapes and transferred them onto a single-crystal diamond. As illustrated in Fig. \ref{fig.1}a, the hBN flakes containing the V${\rm{_{B}^{-}}}$ spin ensemble (see Methods) were precisely placed over NV centers to form a hybrid spin system. The relative positions of the hBN and the NV centers can be further precisely controlled using deterministic transfer methods for 2D materials \cite{zomer2011transfer,castellanos2014deterministic}. In this system, NV and V${\rm{_{B}^{-}}}$ spins exhibit distinct energy levels \cite{du2024single,gottscholl2020initialization} (Fig. \ref{fig.1}b), enabling individual optically detected magnetic resonance (ODMR) measurements (Fig. \ref{fig.1}d, left). Moreover, their photoluminescence (PL) spectra show only partial overlap (Fig. \ref{fig.1}c, top left), allowing for separate fluorescence readout using different optical filters in the collection path (see Supplementary Note 1). Compared to the PL image obtained without any filter (Fig. \ref{fig.1}c, bottom left), selective filtering effectively distinguishes fluorescence signals from NV and V${\rm{_{B}^{-}}}$ spins (Fig. \ref{fig.1}c, top right and bottom right), facilitating independent spin manipulation of each system while maintaining high readout contrast. As shown in the right panel of Fig. \ref{fig.1}d, the measured Rabi contrasts of NV and V${\rm{_{B}^{-}}}$ spins are comparable to those reported for each spin species in previous studies \cite{du2024single,gottscholl2021room,gao2021high}.  This controlled optical and microwave addressability allows for the respective detection of subsequent spin dynamics without sacrificing the readout contrast \cite{barry2020sensitivity,li2022room} or limiting accessibility to a single system \cite{yin2013optical}.

\subsection*{Hybrid magnetic sensing}\label{subsec2}

\begin{figure}[htbp]
\centering
\includegraphics[width=1\textwidth]{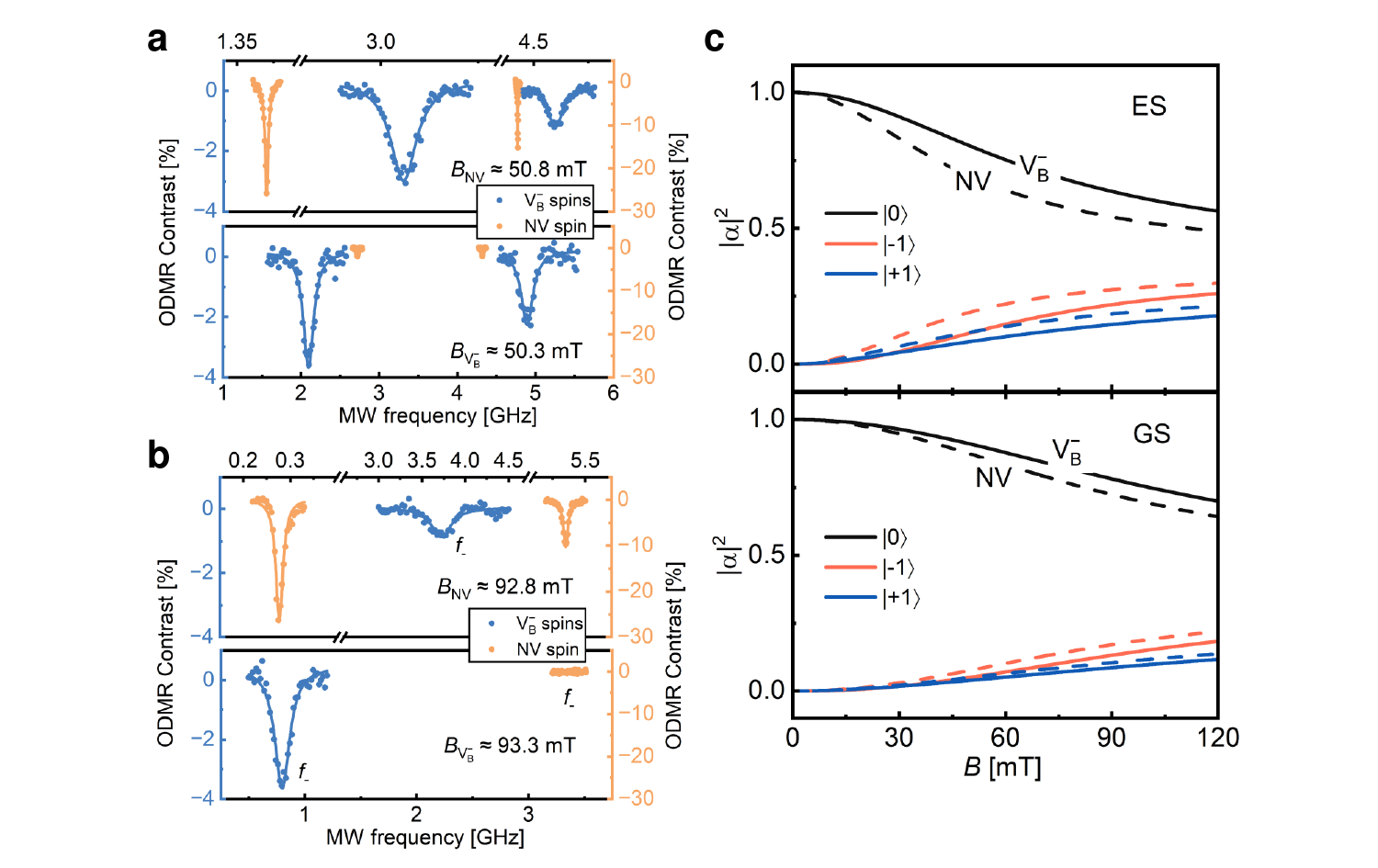}
\caption{\textbf{Hybrid magnetic sensing.} ODMR spectra of NV and V${\rm{_{B}^{-}}}$ spins at magnetic fields of approximately 50 mT \textbf{(a)} and 93 mT \textbf{(b)}, where the magnetic field is aligned along the NV axis ($B$$_{\rm{NV}}$) or the V${\rm{_{B}^{-}}}$ axis ($B_{\rm{V_{B}^{-}}}$), respectively. \textbf{c} Overlap $\left | \alpha  \right | ^{2}$ of each spin level for NV (dashed lines) and V${\rm{_{B}^{-}}}$ spins (solid lines) with $\left | 0  \right \rangle _{z}$ as a function of the magnetic field, where $B$ is at an angle of 54.7$^{\circ}$ to the NV or V${\rm{_{B}^{-}}}$ axis. Upper panel: Calculated results for the ES. Lower panel$:$ Calculated results for the GS.}
\label{fig.2}
\end{figure}

Our hybrid system offers a significant advantage by enabling direct monitoring of variations in spin transition frequency under different external magnetic fields (Fig. \ref{fig.3}a), which can enhance nanoscale magnetic field sensing. One shortcoming of NV sensors is their limited dynamic range for magnetic detection, primarily due to their high sensitivity to the magnetic field direction near a level anti-crossing point \cite{epstein2005anisotropic,tetienne2012magnetic}. As shown in Fig. \ref{fig.2}a and \ref{fig.2}b, the NV center exhibits high ODMR contrast when the magnetic field is aligned along its axis. However, when the magnetic field deviates from the NV axis and aligns with the V${\rm{_{B}^{-}}}$ spin axis, the ODMR contrast of the NV spin significantly decreases (lower panel of Fig. \ref{fig.2}a) or even vanishes (lower panel of Fig. \ref{fig.2}b) \cite{tetienne2012magnetic}. In comparison, the V${\rm{_{B}^{-}}}$ spins maintain high ODMR contrast with the magnetic field either along or off its axis. This difference in sensitivity to the magnetic field primarily arises from the different mixing of spin states in the ground and excited states at the same magnetic field angle \cite{epstein2005anisotropic,baber2021excited,yu2022excited,mu2022excited,mathur2022excited}. Figure \ref{fig.2}c shows $\left | \alpha  \right | ^{2}$, the overlap of $\left | 0  \right \rangle _{z}$ with the spin level of the NV and V${\rm{_{B}^{-}}}$ spins, $\left | m_{s} \right \rangle =\alpha \left | 0  \right \rangle _{z}+\beta \left | -1  \right \rangle _{z}+\gamma \left | +1  \right \rangle _{z}$, where $\left | \beta  \right | ^{2}$ and $\left | \gamma  \right | ^{2}$ are the other respective overlaps, and the subscript $z$ represents the $\left [ 111 \right ] $ axis of diamond or the $c$ axis of hBN. The variation of $\left | \alpha  \right | ^{2}$ with magnetic field $B$, calculated for a magnetic field oriented at an angle of 54.7$^{\circ}$ to their respective axes, indicates that NV spin exhibits a higher probability of spin mixing than V${\rm{_{B}^{-}}}$ spins. Therefore, by integrating thin hBN materials, our hybrid system significantly enhances the dynamic range for nanoscale magnetic sensing, enabling a broader sensitivity window compared to individual systems \cite{beaver2024optimizing}. Furthermore, the hybrid platform offers promising potential for advancements such as entangled quantum sensing \cite{jones2009magnetic,zhou2025entanglement} and nanoscale magnetic resonance using surface spins as reporters \cite{sushkov2014magnetic,schaffry2011proposed}.

\subsection*{Characterization of NV-V${\rm{_{B}^{-}}}$ interaction strength and V${\rm{_{B}^{-}}}$ concentration}\label{subsec3}

\begin{figure}[htbp]
\centering
\includegraphics[width=1\textwidth]{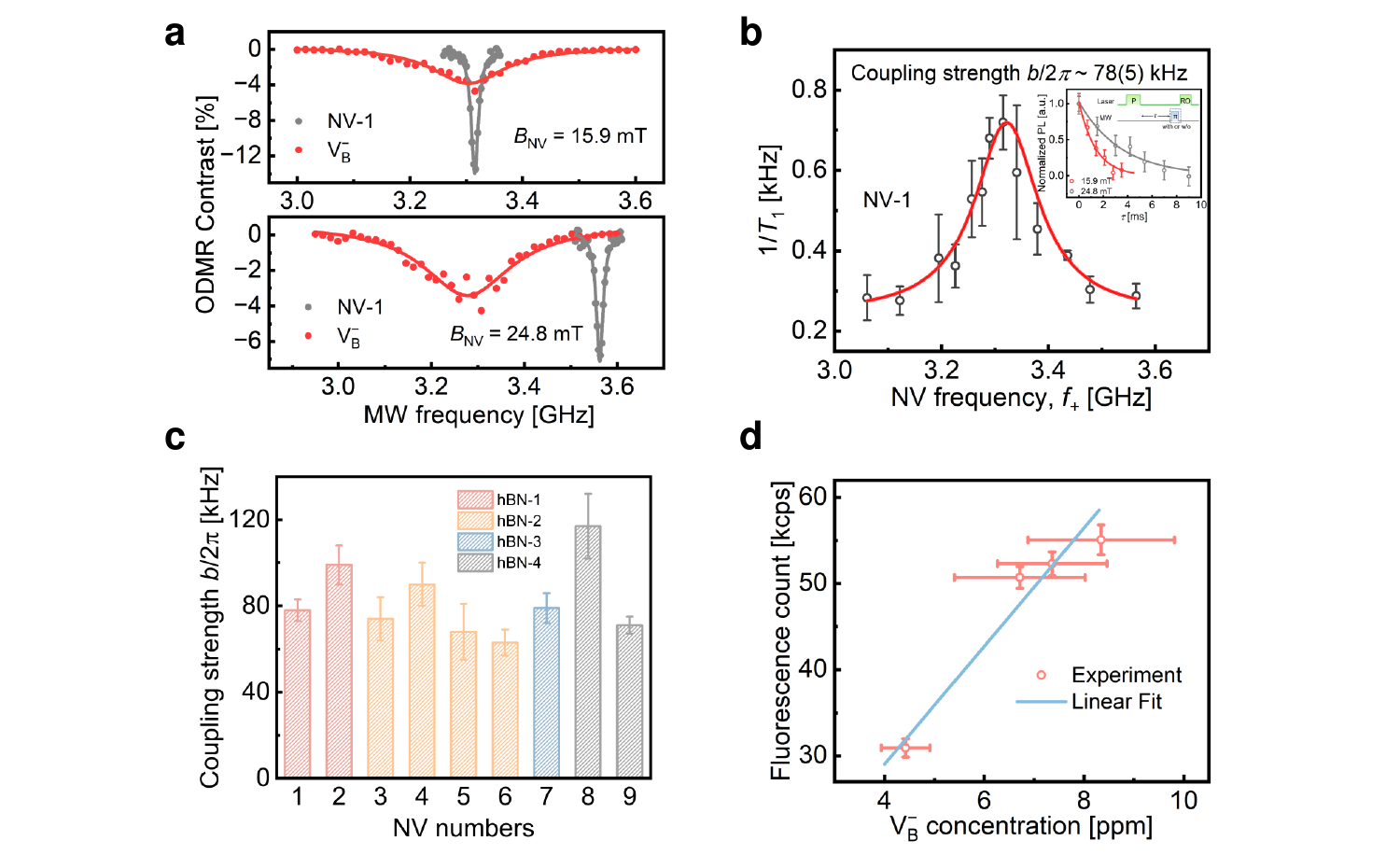}
\caption{\textbf{Characterization of NV-V${\rm{_{B}^{-}}}$ coupling strength and V${\rm{_{B}^{-}}}$ concentration.} \textbf{a} ODMR spectra of NV-1 and V${\rm{_{B}^{-}}}$ spins measured at 15.9 mT (upper panel) and 24.8 mT (lower panel), with the magnetic field aligned along the NV-1 axis. \textbf{b} Longitudinal relaxation rate (1/$T_{1}$) of NV-1 as a function of the transition frequencies, $f_{+}$. The data are fitted using the equation $1/{T_{1}^{\rm{other}}}+{b^{2}}{{\Gamma}/({\Delta^{2}+\Gamma^{2}})}$, with fitted values of $\Gamma/2\uppi$ $\sim$ 160 MHz and $1/{T_{1}^{\rm{other}}}$ $\sim$ 0.24. The inset shows $T_{1}$ measurements of NV-1, where the $T_{1}$ signal is obtained by subtracting two consecutive measurements: one with a $\uppi$ pulse and one without. At 15.9 mT, $T_{1}$ = 1.38(0.13) ms; at 24.8 mT, $T_{1}$ = 3.47(0.38) ms. \textbf{c} Summary of coupling strengths for NV centers under various hBN samples. \textbf{d} Measured fluorescence counts of different hBN samples versus estimated V${\rm{_{B}^{-}}}$ concentration. The data are fitted using a linear function with a slope of 6.86(1.06).}
\label{fig.3}
\end{figure}

Beyond coupling to external magnetic fields, the magnetic dipole-dipole interaction between the NV center and V${\rm{_{B}^{-}}}$ spin ensemble in the hybrid system introduces rich spin dynamics and enables the estimation of the V${\rm{_{B}^{-}}}$ concentration. We first investigate the effect of this interaction on the longitudinal relaxation rate (1/$T_{1}$) of the spin, known as $T_{1}$-based spin relaxometry \cite{hall2016detection,broadway2018quantum}. Under the energy matching condition ($\Delta\ = 2\uppi(f_{{\rm{NV}}}-f_{\rm{V_{B}^{-}}})=0$), energy transfer reduces the $T_{1}$ time of NV-1 to 1.38(0.13) ms. When the energy levels are strongly mismatched (Fig. \ref{fig.3}a, lower panel), $T_{1}$ time increases to 3.47(0.38) ms (see the inset of Fig. \ref{fig.3}b). The detailed dependence on the NV transition frequencies ($f_{+}$) is illustrated in Fig. \ref{fig.3}b, following a Lorentzian distribution \cite{wood2016wide}. In the strong dephasing regime, $b\ll \Gamma $ \cite{hall2016detection} (where $b/2\uppi$ $\sim$ 80 kHz and $\Gamma/2\uppi$ $\sim$ 160 MHz, as detailed later), the relaxation rate can be expressed as $1/{T_{1}}=1/{T_{1}^{\rm{other}}}+{b^{2}}{{\Gamma}/({\Delta^{2}+\Gamma^{2}})}$, where $1/{T_{1}^{\rm{other}}}$ represents the relaxation contributions from other noise, $b$ denotes the interaction strength, and the total defect dephasing rate is $\Gamma = \Gamma_{\rm{NV}}+\Gamma_{\rm{{V_{B}^{-}}}}$.

By fitting the longitudinal relaxation rates shown in Fig. \ref{fig.3}b, we obtain a typical coupling strength ($b/2\uppi$) of 78(5) kHz between NV-1 and the V${\rm{_{B}^{-}}}$ spins, among one of the statistics for various hBN flakes summarized in Fig. \ref{fig.3}c. The coupling strength $b={\gamma_{e}}B_{rms}$, where $B_{rms}$ is the magnetic field component perpendicular to the NV axis, indeed determines the V${\rm{_{B}^{-}}}$ density through \cite{tetienne2013spin,rosskopf2014investigation,myers2014probing}.

\begin{equation}
B_{rms}^2=\sum_iB_{rms,i}^2=\sum_i\frac{\mu_0^2\hbar^2{\gamma_e}^2}{12\uppi}\frac{\rho_i}{d_i^4}\approx\sum_i(3.799\ \mathrm{mT}\ \mathrm{nm}^3)^2\frac{\rho_i}{d_i^4}
\end{equation}
where $B_{rms,i}^2$ denotes the fluctuating magnetic field generated by V${\rm{_{B}^{-}}}$ spins in the $i$-th hBN layer, and $d_{i}$ represents the vertical distance from the NV center to that layer. The NV depth was estimated from DEER spectra of surface spins using a configurationally averaged model (see Supplementary Note 3), while the V${\rm{_{B}^{-}}}$ spin density $\rho_{i}$ in the $i$-th layer is correlated according to the Stopping and Range of Ions in Matter (SRIM) program \cite{ziegler2010srim}. 

We estimated the V${\rm{_{B}^{-}}}$ concentration in hBN-1 samples to be approximately 0.009(1) $\rm{nm}^{-2}$ (see Supplementary Note 2). These concentrations were further verified through the corresponding fluorescence intensity measurement across different hBN samples with consistent laser power (see Supplementary Note 4). As expected, the fluorescence counts exhibited a linear relationship with the estimated concentrations (Fig. \ref{fig.3}d), supporting our coupling strength-derived estimations.

\subsection*{Probing many-body dynamics through a single spin probe}\label{subsec4}

\begin{figure}[htbp]
\centering
\includegraphics[width=1\textwidth]{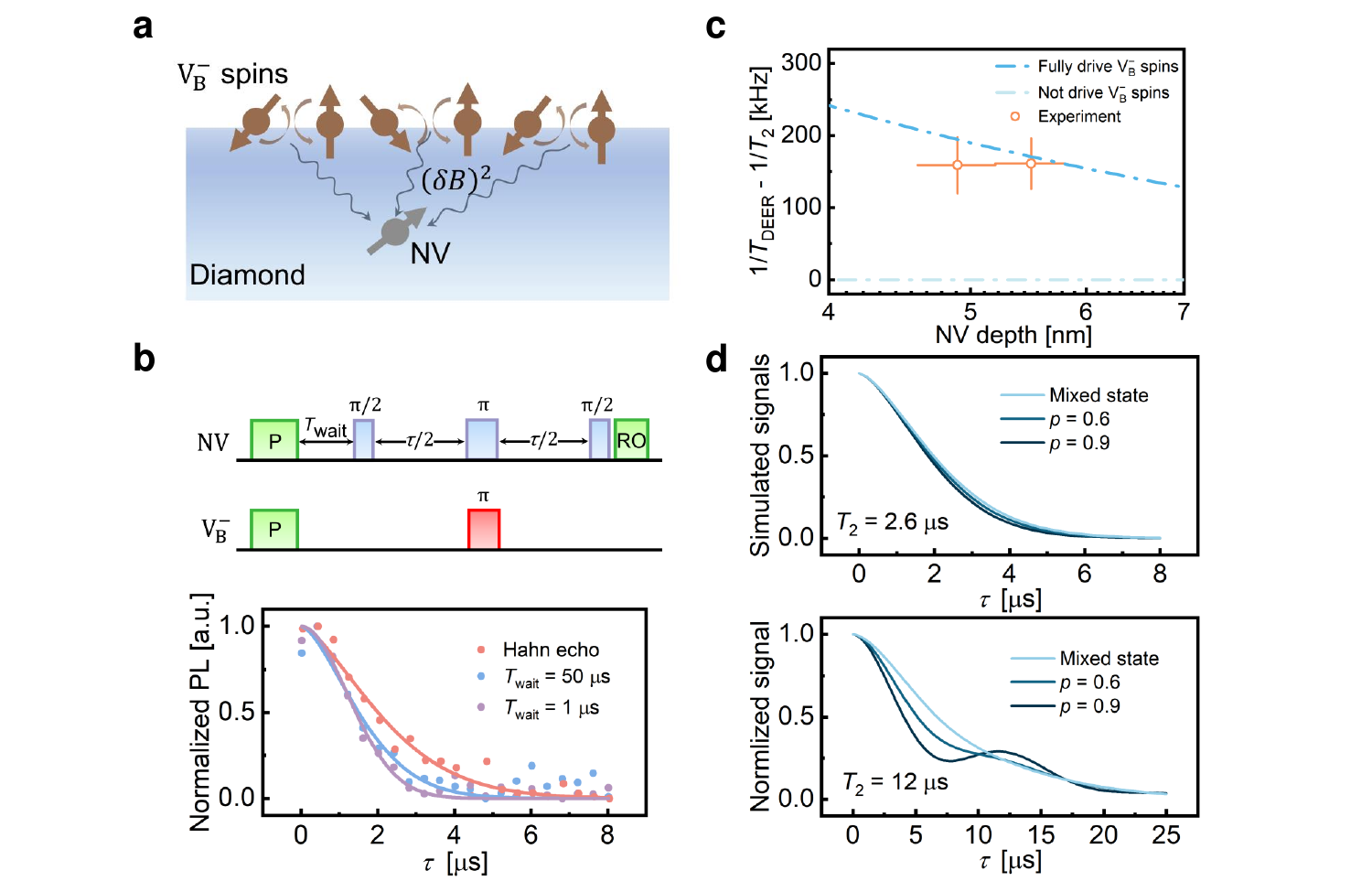}
\caption{\textbf{Probing many-body dynamics through a single spin probe.} \textbf{a} Schematic of fluctuating signal $\left(\delta B \right)^{2}$ measurement using a nearby NV center, where $\left(\delta B \right)^{2}$ is produced by flip-flops between the V${\rm{_{B}^{-}}}$ spins. \textbf{b} Measurements of fluctuating (waiting 50 $\upmu$s) and polarized (waiting 1 $\upmu$s) magnetic field from V${\rm{_{B}^{-}}}$ spins. The upper panel is the DEER pulse sequence, and the lower panel shows the measurement results. The coherence time $T_2$ of the NV center is $\sim$2.6 $\upmu$s, and the measurements were performed at 37.5 mT. \textbf{c} Variation of DEER decay rates with NV depths for fully driving and not driving V${\rm{_{B}^{-}}}$ spins, where the fluorescence-estimated V${\rm{_{B}^{-}}}$ density is $\sim$0.01 $\rm{nm^{-2}}$. The orange dots represent experimental results obtained with a 50 $\upmu$s waiting time. \textbf{d} Simulated DEER signals for varying NV coherence times and polarization degrees of the V${\rm{_{B}^{-}}}$ spins.}
\label{fig.4}
\end{figure}

The high density of V${\rm{_{B}^{-}}}$ spins creates an interacting many-body system, with NV centers serving as effective probes to investigate its dynamics, as illustrated in Fig. \ref{fig.4}a. This hybrid spin system offers excellent controllability, allowing for material engineering to adjust the size of the V${\rm{_{B}^{-}}}$ spin bath and quantum control over spin states, thereby facilitating the study of coherent dynamics amidst many-body noise. To elaborate, we performed double electron-electron resonance (DEER) measurements with two different initial spin states of the V${\rm{_{B}^{-}}}$ spin bath. The V${\rm{_{B}^{-}}}$ spins were prepared in completely mixed states after a 50 $\upmu$s relaxation time, contrasting with polarized spin states from a 1 $\upmu$s wait following optical initialization. As shown in the DEER pulse sequence (Fig. \ref{fig.4}b, top), a Hahn echo sequence was applied to the NV center to filter out unwanted quasi-static noise \cite{dwyer2022probing} while probing the fluctuating magnetic field $\left(\delta B \right)^{2}$  or the polarized magnetic field from V${\rm{_{B}^{-}}}$ spins, with a $\uppi$ pulse applied to the V${\rm{_{B}^{-}}}$ spins. 

For the initial mixed states, the fluctuating magnetic field induced additional decoherence, resulting in a faster decay of the coherence compared to the Hahn echo result without the $\uppi$ pulse (Fig. \ref{fig.4}b, $T_\mathrm{wait}$ = 50 $\upmu$s). By subtracting the decoherence contributions from other sources in the Hahn echo signal \cite{barry2020sensitivity}, we obtained DEER decay rates from V${\rm{_{B}^{-}}}$ spins that fall between the two cases where the V${\rm{_{B}^{-}}}$ spins are fully driven and not driven (Fig. \ref{fig.4}c), primarily due to imperfect spin driving \cite{stepanov2016determination} (see Supplementary Note 5).

For the spin-polarized initial states, the coherence signal exhibited further decay rather than the expected coherent oscillations resulting from the net static magnetic field of the polarized V${\rm{_{B}^{-}}}$ spins (Fig. \ref{fig.4}b, $T_\mathrm{wait}$ = 1 $\upmu$s and Supplementary Figure 13). This behavior can be understood by noting the competitive effects of coherent evolution and rapid decoherence. As shown in Fig. \ref{fig.4}d, the short coherence time of 2.6 $\upmu$s (observed in Fig. \ref{fig.4}b) leads to faster decay in the numerical simulations of DEER signals as the polarizations ($p$) of the V${\rm{_{B}^{-}}}$ spins increase (see Supplementary Note 6). This occurs because the faster oscillation induced by the increased static magnetic field with higher polarizations is masked by the limited coherence time. Conversely, with a longer spin coherence time of 12 $\upmu$s, the DEER signals display the anticipated coherent oscillations.

\section*{Discussion}\label{sec3}

In this work, we demonstrate coherent control of a hybrid spin system that operates at room temperature within easily integrated 2D–3D heterostructures. By combining V${\rm{_{B}^{-}}}$ spins with a single NV center, this system enables enhanced magnetic sensing, estimation of the V${\rm{_{B}^{-}}}$ spin concentration, and the exploration of its rich many-body dynamics. The controllability of the spin systems opens up new opportunities for applications, such as using V${\rm{_{B}^{-}}}$ spins as reporters for nanoscale magnetic resonance \cite{sushkov2014magnetic,schaffry2011proposed,zhang2023reporter}. Furthermore, our approach can be adapted to hybrid systems incorporating other defects in hBN with excellent coherence properties \cite{chejanovsky2021single,mendelson2021identifying,stern2022room,guo2023coherent,stern2024quantum}. By combining precise engineering techniques \cite{wang2022self}, this 2D-3D heterostructure platform provides a promising route for advancing quantum technologies.

\section*{Methods}\label{sec4}

\subsection*{Sample preparation}

The hBN samples were purchased from HQ Graphene. Sample hBN-1, hBN-2 and hBN-5 were implanted with nitrogen ions at a dose of $5\times 10^{13}\:\rm{ions/cm^2}$, whereas hBN-3 and hBN-4 were implanted with $1\times 10^{14}\:\rm{ions/cm^2}$. All hBN samples were implanted with $\rm{^{14}N^{+}}$ at an energy of 2.5 keV. Three (100)-oriented diamond samples, labeled S1, S2, and S3, were used in this experiment. hBN-1 and hBN-2 were transferred onto sample S1, hBN-3 and hBN-4 onto sample S2, and hBN-5 onto sample S3. Sample S1 and S2, with dimensions of $2\rm{mm}\times 2\rm{mm}\times 0.5\rm{mm}$, were used for $T_1$ measurements, while sample S3, with dimensions of $1\rm{mm}\times 1\rm{mm}\times 0.05\rm{mm}$, was used for DEER measurements. All diamond samples were treated with reactive ion etching and high-temperature annealing prior to ion implantation, as described in reference \cite{sangtawesin2019origins}, to enhance the stability of the shallow NV centers. The NV centers in all samples were created by $\rm{^{15}{N_2}^{+}}$ ion implantation at a dosage of $1\times \rm{10^9 /cm^2}$ and an energy of 5 keV. After ion implantation, the samples were annealed in high vacuum ($\sim$$10^{-8}$ Torr) at 1000 $^\circ$C for 4 hours. 

\subsection*{Experimental setup}

A home-built confocal microscopy system combined with a MW system was used for the spin manipulation experiments (see Supplementary Note 1). A 532 nm laser and an objective lens (Olympus, LUCPLFLN 60X N.A. = 0.7) were used to excite and collect fluorescence from NV and V${\rm{_{B}^{-}}}$ spins. The laser was modulated by an acousto-optic modulator (AOM). Fluorescence from NV and V${\rm{_{B}^{-}}}$ spins was separately filtered using a 785 nm long-pass filter (Semrock, BLP01-785R-25) and a 758 nm short-pass filter (Semrock, FF01-758/SP-25). Eventually, the fluorescence split into two equal-intensity beams by a D-shaped mirror and were collected by two avalanche photodiode detectors (Excelitas SPCM-AQRH-24). The microwave driving field was generated by mixing the output from two microwave sources (Stanford Research SG386 and SG396). The microwave signal was amplified by a microwave amplifier (Mini-Circuits ZHL-16W-43+) and delivered through a 20-$\upmu$m diameter copper wire (for $T_1$ measurements to characterize interaction strengths, Supplementary Figure 4a) or through a gold film microwave stripline (for Rabi and DEER measurements, Fig. \ref{fig.2}a). The microwave stripline (10-$\upmu$m wide) was fabricated lithographically over the $\rm{SiO_2/Si}$ substrate by e-beam evaporation (10 nm Ti/100 nm Au/20 nm Ti).


\section*{Data availability}
The datasets used and/or analyzed during the current study available from the corresponding author upon reasonable request.

\section*{Code availability}
All codes that support the findings of this study are available from the corresponding author upon reasonable request.

\bibliography{references}

\section*{Acknowledgements}
The fabrication of diamond and hBN samples, as well as the gold film microwave stripline, was partially performed at the USTC Center for Micro and Nanoscale Research and Fabrication, and the authors particularly thank Cunliang Xin, and Hongfang Zuo for their assistance in the fabrication process. This work is supported by the National Natural Science Foundation of China (Grants No.T2325023, 92265204, 12474500, 12104447), the Innovation Program for Quantum Science and Technology (Grant No. 2021ZD0302200).

\section*{Author contributions}
Y.W. supervised the project. Y.W. and M.W designed the experiments. M.W., Z.L., Y.G., Y.H., F.S. and P.W. built the experimental setups. H.S. and P.Y. performed the experiments. K.X. and W.L. conducted the photoluminescence measurements. H.S., P.Y., X.Y., and M.W. prepared the sample. Z.X., H.S. and Y.P. performed the calculations and simulations. H.S., Y.W., P.Y. and Z.X. wrote the manuscript. All authors discussed the results and commented on the paper.

\section*{Competing interests}
The authors declare no competing interests.

\end{document}